\documentclass[aps,prb,twocolumn,superscriptaddress]{revtex4}
\usepackage{graphicx}
\usepackage{amsmath}
\usepackage{amssymb}
\usepackage{graphics}
\usepackage{bm}
\usepackage{dcolumn}
\begin{document}
\title[]{Circular dichroism simulated spectra of chiral gold nanoclusters:
         A dipole approximation}

 \author{Carlos E. Rom\'an-Vel\'azquez}
 \affiliation{Centro de Investigaci\'on en Ciencia Aplicada y
 Tecnolog\'{\i}a Avanzada, Instituto Polit\'ecnico Nacional,
 Av. Legaria 694, 11500 M\'exico, D.F., M\'exico}
 \author{Cecilia Noguez}
 \email[Author to whom all correspondence should be addressed.
  E-mail:]{cecilia@fisica.unam.mx}
 \affiliation{Instituto de F\'{\i}sica, Universidad Nacional
 Aut\'onoma de M\'exico, Apartado Postal 20-364,
 01000  M\'exico, D.F., M\'exico}
 \author{Ignacio L. Garz\'on}
 \affiliation{Instituto de F\'{\i}sica, Universidad Nacional
 Aut\'onoma de M\'exico, Apartado Postal 20-364,
 01000  M\'exico, D.F., M\'exico}
 \date{\today}


\begin{abstract}
Circular dichroism (CD) spectra of chiral bare and
thiol-passivated gold nanoclusters
have been calculated within the  dipole
approximation. The calculated CD spectra show features that allow us to
distinguish between clusters with different indexes of chirality.
The main factor responsible of the differences in the CD lineshapes
is the distribution of interatomic distances that
characterize the chiral cluster geometry.
These results
provide theoretical support for the quantification of chirality
and its measurement, using the CD lineshapes of chiral metal nanoclusters.
\end{abstract}

\maketitle

Recent theoretical studies on the structural properties of gold clusters have shown that the lowest-energy (most stable) isomers of bare Au$_{28}$ and
Au$_{55}$, and thiol-passivated Au$_{28}$(SCH$_3$)$_{16}$ and Au$_{38}$(SCH$_3$)$_{24}$  clusters correspond to chiral nanostructures \cite{GarzonPRB,GarzonEPJ}. These results provided theoretical support for the existence of chirality in metal clusters, suggested by the intense optical activity measured in the metal-based electronic transitions of size-separated glutathione-passivated gold nanoclusters in the size range of 20-40 atoms \cite{Schaaff}.

Another theoretical prediction, based on quantifying chirality through the Hausdorff chirality measure (HCM) \cite{GarzonPRB,GarzonEPJ,BudaJACS}, indicates that the strong structural distortion in a gold cluster upon thiol passivation could induce chirality in an achiral bare cluster or increase the index of chirality in an already bare chiral cluster \cite{GarzonPRB,GarzonEPJ}. To confirm this prediction further experimental studies on chiral bare and passivated clusters are expected to be performed in the near future using, for example, circular dichroism (CD) spectroscopy \cite{Peacock}.

To explore the capability of the CD technique to detect different
indexes of chirality existing in bare and passivated gold nanoclusters it would be useful to have a theoretical estimation of their CD lineshapes.
Quantum-mechanical calculations of CD spectra for small and moderately sized organic molecules had been performed \cite{Ziegler}, however, this type of approach is still inappropriate for passivated gold nanoclusters due to the huge computational effort that would be involved \cite{Ziegler}. In this work, we present results on simulated CD spectra of chiral gold nanoclusters obtained through a dipole approximation \cite{purcell,draine}. This study allows
to gain insights on the variations of the CD lineshape due to different indexes of chirality existing in gold clusters. It is expected that this information would be
useful to motivate further experimental studies that confirm not only the existence of chirality in gold nanoclusters, but also to correlate distinct features of the CD spectra with the different indexes of chirality \cite{GarzonPRB,GarzonEPJ}.

In this work  the nanoclusters of interest are composed by
$N$ gold atoms. We suppose that each one of these atoms
is represented by a polarizable point dipole located at the
position of the atom. We assume that the dipole located at
$\mathbf{r}_{i}, \, \, i=1,2, \dots ,N$, is characterized by a
polarizability $\boldsymbol{\alpha}_{i} (\omega)$, where $\omega$
denotes the angular frequency. The cluster is excited by a
circularly polarized incident wave
$\mathbf{E}_{\mathrm{inc} }=
E_{\mathrm{0}} (\hat{\bf x} \pm \imath \hat{\bf y})
e^{i\mathbf{k\cdot r} -i\omega t}
$, where $t$ denotes time, $\mathbf{k}$ is the wave-vector,
whose magnitude is given by $k=\omega /c=2\pi /\lambda $,
$c$ is the speed of light, and $\lambda $ is the wavelength
of the incident light. Here $\hat{\bf x}$ and $\hat{\bf y}$
denotes the unitary vector along any of two orthogonal directions, and the sign $+$ ($-$) corresponds to right- (left-) circularly polarized light.

Each dipole of the system is subject to a total electric field
that can be divided into two contributions: (i) the incident
radiation field, plus (ii) the radiation field resulting from
all of the other induced dipoles. The sum of both fields is
the so called local field and is given by
\begin{equation}
\mathbf{E}_{i,\mathrm{loc}} = \mathbf{E}_{i,\mathrm{inc}} +
\mathbf{E}_{i,\mathrm{dip}} = \mathbf{E}_{i,\mathrm{inc}} -
\sum_{i\neq j} \mathbb{T}_{ij} \cdot \mathbf{p}_{j},
\end{equation}
where $\mathbf{p}_{i}$ is the dipole moment of the atom
located at $\mathbf{r}_{i}$, and $\mathbb{T}_{ij}$ is an
off-diagonal interaction matrix with $3\times 3$ matrices
as elements, such that
\begin{eqnarray}
\mathbb{T}_{ij} \cdot \mathbf{p}_{j} &=& \frac{e^{ikr_{ij}}}
{r_{ij}^{3}} \Bigg\{k^{2}
\mathbf{r}_{ij} \times (\mathbf{r}_{ij}\times \mathbf{p}_{j}) \\
&+& \frac{(1-ikr_{ij})} {r_{ij}^{2}}\left[ r_{ij}^{2}
\mathbf{p}_{j}-3 \mathbf{r}_{ij} (\mathbf{r}_{ij}\cdot
\mathbf{p}_{j})\right] \Bigg\}.  \notag
\end{eqnarray}
Here $r_{ij}=|\mathbf{r}_{i}-\mathbf{r}_{j}|,$
and $\mathbf{r}_{ij} =\mathbf{r}_{i} - \mathbf{r}_{j}$.
On the other hand, the induced dipole moment at each atom is given by
$\mathbf{p}_{i}=\boldsymbol{\alpha}_{i}\cdot \mathbf{E}_{i,\mathrm{loc}}$, such that $3N$-coupled complex linear equations are obtained. These equations can  be rewritten as
\begin{equation}
\sum_{j}\left[ \frac{1} {\boldsymbol{\alpha}_{j}} \delta_{ij} +
 \mathbb{T}_{ij} (1 - \delta_{ij}) \right] \cdot \mathbf{p}_{j} = 
\mathbb{M}_{ij} \cdot \mathbf{p}_{j} =  \mathbf{E}_{i,\mathrm{inc}} \, .
 \label{pes}\end{equation}
The matrix $\mathbb{M}$ is composed by a diagonal part given
by $\frac{1} {\boldsymbol{\alpha}_{j}} \delta_{ij}$ and by an
off-diagonal part given by $\mathbb{T}_{ij}$.
The diagonal part is only related with the material properties
of each atom, while the off-diagonal part only depends on
the geometrical properties of the system.

Once we solve the complex-linear equations shown in Eq. (\ref{pes}),
the dipole moment on each atom in the cluster can be determined.
Then, we can find the
extinction  cross section of the whole nanoparticle,
in terms of the dipole moments, as~\cite{purcell,draine}
\begin{equation}
C_{\mathrm{ext}} =\frac{4\pi k}{{E}_{0}^{2}} \sum_{i=1}^{N}
\mathrm{Im} (\mathbf{E}_{i,\mathrm{inc}}^{\ast }\cdot \mathbf{p}_{i}) \,
 ,
\end{equation}
where $\ast $ means complex conjugate.
The extinction efficiency is defined as $Q_{\mathrm{ext}}=C_{\mathrm{ext}}/A
$, where $A = 4 \pi d^2$, and $d$ is the maximum distance
between any two atoms in the nanocluster. Finally, the circular dichroism
spectra is obtained by subtracting the averaged extinction
efficiencies for left- and right-circularly polarized light,
\begin{equation}
{\mathrm CD} = <Q_{\mathrm{ext}}^R>_S - <Q_{\mathrm{ext}}^L>_S , \label{cd}
\end{equation}
where $ < \cdots >_S$ indicates an average of the extinction
efficiency, calculated over $S$ different angular incident
directions of the electric field~\cite{prom}.

In this work, we assume that the polarizability for each atom $\alpha_i$
is the same for all the atoms in the cluster ($\alpha_i$ = $\alpha$), and is
given by
the well-known Clausius-Mossotti relation
\begin{equation}
\alpha=\left(\frac{R}{3}\right)^3 \frac{(\epsilon - 1)}{(\epsilon+2)},
\end{equation}
where $\epsilon$ is the macroscopic dielectric function of the nanoparticle,
and  $R$ is the ionic radii of the atom.
The dielectric function is taken from the measurements by Johnson and
Christy~\cite{JandC} for bulk gold.
This approximation has been used before
providing  good qualitative agreement between calculated optical
properties and experimental data on gold and silver nanoparticles \cite{Roman}.
On the other hand, since chirality is a geometrical property,
the CD spectrum will be more sensitive to the second term inside
the bracket in Eq.~(\ref{pes}), which depends on the cluster geometry,
than to the first one that is related with the atomic polarizability.
Therefore, in the present calculations the simulated CD spectra
will include the geometrical
information about the chiral clusters, despite
the polarizability is incorporated in an approximated form.

The lowest-energy structures of the bare Au$_{28}$ and Au$_{38}$,
and thiol passivated Au$_{28}$(SCH$_{3}$)$_{16}$ and
Au$_{38}$(SCH$_{3}$)$_{24}$ clusters had been obtained using density
functional theory (DFT) within the generalized gradient approximation
(GGA) \cite{GarzonPRB,GarzonEPJ}. Both, the bare and thiol passivated
most stable 28-atom gold clusters were found to be chiral, but a higher index
of chirality was obtained for the passivated cluster.
This lowest-energy passivated cluster was found
upon relaxation of the bare chiral cluster, interacting with 16 \, SCH$_{3}$
molecules placed close to the three-atom hollow sites on the
metal cluster surface \cite{GarzonPRB,GarzonEPJ}.
The most stable thiol-passivated Au$_{38}$(SCH$_{3}$)$_{24}$ cluster
was also found chiral \cite{GarzonPRB,GarzonEPJ},
but it was obtained from the relaxation of
the achiral Au$_{38}$ cluster with $O_h$ symmetry interacting with
24 \, SCH$_{3}$ molecules placed on the three-atom hollow sites of the
(111) faces of the bare cluster \cite{GarzonPRL}.
The Au$_{38}$ isomer with $O_h$ symmetry corresponds to a local
minimum of the bare cluster according to DFT-GGA
calculations \cite{Soler}, while the lowest-energy structure of the bare
Au$_{38}$ was found with $C_s$ symmetry \cite{Soler}, but upon passivation,
it transforms into a higher-energy passivated isomer \cite{GarzonPRL}.

\begin{figure}[htb]
\caption{ \label{Fig1} Circular dichroism spectra of Au$_{38}$ clusters.
(a) Before, with $O_h$ symmetry, (dotted line) and after, with $C_1$ symmetry,
(full line) passivation with
24 methylthiol molecules. (b) Lowest-energy isomer, with $C_s$ symmetry,
(full line) and the same cluster geometry but taking out 10 atoms
that are placed on the plane of symmetry (dotted line).
}
\end{figure}
In this work, we use the relaxed Cartesian coordinates of the
clusters mentioned above to calculate their CD spectra. Figure 1(a)
shows the CD lineshape calculated using Eq. (\ref{cd}) for the Au$_{38}$
cluster before (dotted line)  and after (full line) relaxation
with the 24 methylthiol molecules.
The zero value of the CD spectra before relaxation indicates that the
left- and right-circularly polarized light is equally absorbed by the gold
cluster since its geometrical structure corresponds to a truncated
octahedral structure with  $O_h$ symmetry. Since the geometry
of this cluster is highly symmetric,
its CD spectrum should be zero in agreement with our result.
On the other hand, the CD spectrum of the passivated cluster
shows a non-zero lineshape due to the different absorption of the
circularly polarized light. This behavior is expected since the
passivated Au$_{38}$ has been found chiral using the geometrical
approach described in Refs. \onlinecite{GarzonPRB,GarzonEPJ}.

Figure 1(b) shows an interesting result on the relation between chirality
and the lineshape of the CD spectrum. The full line corresponds to the
calculated CD spectrum of the lowest-energy structure of the bare Au$_{38}$.
The structure of this isomer has one plane of symmetry ($C_s$ symmetry)
and therefore is achiral \cite{GarzonPRB,GarzonEPJ}. However, as Fig. 1(b)
shows, the CD spectrum is non-zero. The reason of such behavior is
coming from the atomic array existing on the plane of symmetry.
This plane contains 10 atoms that are not in symmetric positions
but forming a chiral two-dimensional array. This subset of atoms
generate the non-zero signal of the CD spectrum. To confirm this effect,
the CD lineshape of the 38-atom  cluster without
including the atoms on the plane of symmetry was calculated (dashed
line). Since in this case there is
a complete reflection symmetry in the cluster, its CD spectrum is
negligible.

\begin{figure}[htb]
\caption{ \label{Fig2} Circular dichroism spectra of Au$_{28}$ clusters.
(a) Before, with $C_1$ symmetry, and (b) after, with $C_1$ symmetry,
passivation with
16 methylthiol molecules.
}
\end{figure}
Figures 2(a) and 2(b) display the CD spectra of the Au$_{28}$
cluster before (bare)
and after thiol relaxation (passivated), respectively. Both
clusters are chiral according to the geometrical analysis, based
on the HCM, described
in Refs. \onlinecite{GarzonPRB,GarzonEPJ}. Therefore, it is expected
that they have a non-zero signal in their CD spectra. However, Fig. 2 also
shows that indeed there exist notorious differences between the CD lineshapes
of these clusters. These differences could account for the distinct
indexes of chirality existing in these clusters as was calculated
before using the HCM \cite{GarzonPRB,GarzonEPJ}. In other words, Fig. 2
shows that clusters with different values of HCM can be distinguished
through the lineshape of their CD spectra.

To understand the origin of the differences in the CD
spectra of clusters with different values of their indexes of chirality,
the extinction efficiencies, using
right-circularly polarized light,
are shown in Figs. 3 (a) and (b) for the bare and passivated Au$_{28}$
clusters, respectively. Since the CD spectrum is calculated from the
difference of this quantity with the one obtained using left-circularly
polarized light, it is interesting to analyze the most important
features of its lineshape. From Figs. 3 (a) and (b) it is evident that
the clusters with different HCM values also generate distinguishable
extinction efficiencies.
By analyzing Eqs. (\ref{pes} -- \ref{cd}), it is found that
the cluster geometry, through the distribution of interatomic distances
$r_{ij}$, is the main factor that generate the features (peaks and  widths in the lineshape) of the extinction efficiency.
\begin{figure}[htb]
\caption{ \label{Fig3} Extinction efficiency using
right-circularly polarized light of Au$_{28}$ clusters.
(a) Before, with $C_1$ symmetry, and (b) after, with $C_1$ symmetry,
passivation with
16 methylthiol molecules. The insets show the geometry of the
bare and passivated clusters.
}
\end{figure}

The effect of the distribution of interatomic distances in the
behavior of the extinction efficiency is more evident by
comparing this quantity for the 38-atom cluster,
before and after relaxation with the thiol molecules. Figure 4 shows
the extinction efficiency using right-circularly polarized light
for the bare Au$_{38}$ isomer with $O_h$ symmetry (upper panel)
and for the passivated chiral cluster (bottom panel).
The insets in Fig. 4 display the distribution of interatomic
distances of the corresponding clusters. From these figures it can
be noticed that the extinction efficiency of the chiral passivated
cluster (bottom panel) has more structure and shows a broader lineshape
than the achiral cluster (top panel). This difference can be correlated
with the different behavior of the distribution of interatomic distances
that indicate a discrete variation of distribution for the achiral
$O_h$ cluster, whereas the chiral one shows a much broader
and continuous distribution
of interatomic distances.
\begin{figure}[htb]
\caption{ \label{Fig4} Extinction efficiency using
right-circularly polarized light of Au$_{38}$ clusters.
(a) Before, with $O_h$ symmetry, and (b) after, with $C_1$ symmetry,
passivation with 24 methylthiol molecules.
The insets show the distribution of interatomic
distances and the geometries of the bare and passivated clusters.
}
\end{figure}

The results presented in this work indicate that indeed chiral
gold clusters with different indexes of chirality can be distinguished
by the lineshape of their CD spectra. It was also found that within
the  dipole approximation,
the origin of the distinct features of the CD lineshapes
is related to the differences in the distribution of interatomic
distances. Therefore, CD spectra are able to detect the geometrical
features that characterize clusters with different indexes of
chirality. These geometrical differences will also be important in
quantum mechanical calculations of the CD lineshapes.
It is encouraging that despite the approximations included in the calculation
of the simulated CD spectra, their lineshapes show similar features
as those measured for glutathione-passivated gold
nanoclusters \cite{Schaaff}.
In particular, there is a qualitative agreement between the calculated and
experimental CD spectra in the energy range (2-4 eV), where the spectra shows
the highest intensity .

In summary, CD simulated spectra are appropriate for the characterization
of clusters with different chirality and provide useful information for
further studies using a quantum mechanical approach, and for
experimental studies on the quantification of chirality in
metal nanoclusters.

\acknowledgments
We thank Ivan O. Sosa for useful discussions at the initial stage of this work. This work has been partly supported by DGAPA-UNAM grants No.~IN104201 and IN104402, and by CONACyT grant~36651-E. 
\appendix


\begin{thebibliography}{99}

\bibitem{GarzonPRB} I.L. Garz\'on, J.A. Reyes-Nava,
J.I. Rodr\'{\i}guez-Hern\'andez,
I. Sigal, M.R. Beltr\'an, K. Michaelian,
Phys. Rev. B
\textbf{66}, (2002) 073403.

\bibitem{GarzonEPJ} I.L. Garz\'on, M.R. Beltr\'an,
G. Gonz\'alez, I. Gut\'{\i}errez-Gonz\'alez, K. Michaelian,
J.A. Reyes-Nava,
J.I. Rodr\'{\i}guez-Hern\'andez,
Eur. Phys. J. D
\textbf{xx}, (2003) In press.

\bibitem{Schaaff}
G.T. Schaaff, R.L. Whetten,
J. Phys. Chem. B
\textbf{104}, (2000) 2630.

\bibitem{BudaJACS}
A.B. Buda, K. Mislow,
J. Am. Chem. Soc.
\textbf{114}, (1992) 6006.

\bibitem{Peacock}
R.D. Peacock, B. Stewart,
J. Phys. Chem. B
\textbf{105}, (2001) 351.

\bibitem{Ziegler}
J. Autschbach, T. Ziegler, S.J.A. van Gisbergen, E.J. Baerends,
J. Chem. Phys.
\textbf{116}, (2002) 6930.

\bibitem{purcell} 
E.M. Purcell and C.R. Pennypacker, 
Astrophys. J.
\textbf{186}, (1973) 705.

\bibitem{draine}  
B. T. Draine,  
Astrophys. J., 
\textbf{333}, (1998) 848.

\bibitem{prom} 
Tipically, we employ a minimum of 13,500 different directions.

\bibitem{JandC} 
P. B. Johnson and R. W. Christy, 
Phys. Rev. B
\textbf{6}, (1972) 4370.

\bibitem{Roman} 
I.O. Sosa, C. Noguez, and R.G. Barrera, 
J. Phys. Chem. B; 2003; 
ASAP Web Release Date: 07-Jun-2003; (Article) DOI: 10.1021/jp0274076.

\bibitem{GarzonPRL} I.L. Garz\'on, C. Rovira, K. Michaelian,
M.R. Beltr\'an, P. Ordej\'on, J. Junquera, D. S\'anchez-Portal,
E. Artacho, and J.M. Soler,
Phys. Rev. Lett.
\textbf{85}, (2000) 5250.

\bibitem{Soler} J.M. Soler, M.R. Beltr\'an, K. Michaelian, I.L. Garz\'on,
P. Ordej\'on, D. S\'anchez-Portal, and
E. Artacho.
Phys. Rev. B
\textbf{61}, (2000) 5771.


\end{thebibliography}
\end{document}